\documentclass[12pt,a4paper]{article}
\usepackage[T1]{fontenc}
\usepackage[latin1]{inputenc}
\usepackage[english]{babel}
\usepackage{latexsym}
\usepackage{setspace}
\usepackage[dvips,usenames]{color}

\def\be{\begin{equation}}
\def\fin{\end{equation}}
\def\disp{\displaystyle}

\defþ{\theta}

\def\T{{\sf T\kern-.45em T}}
\def\C{\kern.1em{\raise.47ex\hbox{$\scriptscriptstyle |$}}
             \kern-.40em{\sf C}}

\def\hfl{\disp\mathop{\hbox to 10mm{\rightarrowfill}}}

\usepackage{graphicx,amssymb,amsmath,apalike}

\title{Kinetics of target site localization of a protein on DNA: a stochastic approach}
\author{M. Coppey*, O. Bénichou$^\#$, R. Voituriez* and M. Moreau* \\ \small{
*Laboratoire de Physique Théorique des Liquides, Université Pierre et
Marie Curie,} \\ \small{ case courrier 121, 4 Place Jussieu -- 75252 Paris cedex
05, France} \\  \small{ $^\#$Laboratoire de Physique de la Matière Condensée,
Collège de France, } \\ \small{11 Place M.Berthelot, 75252 Paris Cedex 05, France}}

\begin{document}

\bibliographystyle{plain}

\maketitle

\doublespacing

\begin{abstract}

It is widely recognized that the cleaving rate of a restriction
enzyme on target DNA sequences is several orders of magnitude
faster than the maximal one calculated from the diffusion--limited
theory. It was therefore commonly assumed that the target site
interaction of a restriction enzyme with DNA has to occur via two
steps: one--dimensional diffusion along a DNA segment, and
long--range jumps coming from association/dissociation events. We
propose here a stochastic model for this reaction which comprises a series of 1D
diffusions of a restriction enzyme on non-specific DNA sequences
interrupted by 3D excursions in the solution until the target
sequence is reached. This model provides an optimal finding
strategy which explains the fast association rate. Modeling the
excursions by uncorrelated random jumps, we recover
the expression of the mean
time required for target site association to occur given by Berg \& al. \cite{berg81}, and we explicitly give
several
physical quantities describing the stochastic pathway of the
enzyme. For competitive target sites we calculate two
quantities: processivity and preference. By
comparing these theoretical expressions to recent experimental
data obtained for \textit{Eco}RV--DNA interaction, we
 quantify: i) the mean residence time per binding event of
\textit{Eco}RV on DNA for a representative 1D diffusion coefficient, ii) the average lengths of DNA scanned
during the 1D diffusion (during one binding event and during the
overall process), iii) the mean time and the mean number
of visits needed to go from one target site to the other. Further, we
evaluate the dynamics of DNA cleavage with regard to the
probability for the restriction enzyme to perform another
1D diffusion on the same DNA substrate following a 3D excursion.

\end{abstract}

\noindent {\small Keywords: stochastic, diffusion, restriction enzyme,
target site, DNA}

\vspace{0.2in}

\newpage

{\Large \textbf{Introduction}}

Genetic events often depend on the interaction of a restriction
enzyme with a target DNA sequence. Indeed, the restriction enzyme
has first to find this sequence on DNA. This mechanism has long
remained mysterious. The simplest model considers this mechanism
as a reaction between two point--like entities, the restriction
enzyme and its target DNA sequence, in a solute volume. However,
kinetic measurements of reactivity show that the reaction occurs
at an extraordinarily rapid rate, far above the three-dimensional
diffusion limit rate! \cite{richter74,riggs70}. To account for this, it was
proposed that the reaction occurs via a "facilitated" diffusion
process \cite{vonhippel89}. The restriction enzyme first binds to
DNA on a non--specific site, then performs a one--dimensional
random walk until it reaches the target DNA sequence. Indeed, it is
by scanning the DNA and not by diffusing in a 3D volume that the
restriction enzyme reaches its target site sequence. However, results from experiments
\cite{szczelkun96} using two interlinked rings of DNA (plasmid,
each containing a target site for the restriction enzyme
\textit{Eco}RV) rule out this possibility: the mechanism of target
site localization does  not involve a unique 1D diffusion along
DNA. If it were the case, the \textit{Eco}RV enzyme would cleave
the DNA of only one of the two rings, as opposed to what is
observed. Moreover, it is expected that molecular crowding of
in--vivo situations must hinder any long 1D scanning process of
the DNA \cite{wenner99}.

To account for the fast association rate, several strategies have
been proposed and modeled from experimental data \cite{berg81,vonhippel89,winter81}. Four major translocation
processes were identified (we recall that translocation is the
overall process by which a protein goes from one DNA
sequence to another). The first, the "sliding" process, corresponds to the
pure one--dimensional diffusion as discussed above. The second,
the " intersegmental transfer" \cite{milsom01}, involves dimer
proteins having two binding sites. The restriction enzyme bound on
DNA at the first site binds its second site to a remote DNA
sequence and then dissociates from the first one. The two other
translocation processes are induced by several
dissociation-reassociation events. According to the rebinding of
the enzyme either near the departure site or to an uncorrelated
site, the translocation process is called "hopping" or "jumping"
\cite{halford02}. Which of these translocation processes or which
combination of them describes the mechanism of target site localization
on DNA is still an open question.

Understanding the translocation process is of great importance as
it governs the kinetics of genetic events \cite{mistelli01}.
Several experimental investigations were carried out in order to
elucidate the pathway followed by a restriction
enzyme to reach a single target site. Some of them
 quantify the rate of cleavage reactions,
by varying the length of the DNA strand (for a review, see
\cite{shimamoto99}) or the salt concentration
\cite{winter81,lohman96} which affects the binding properties of
DNA-affine proteins on non-specific sequences. These experimental results
allows one to reject the possibility of a unique translocation
process, but can not fully describe the structure of the combined
process. Berg \& al. \cite{berg81} had proposed a theoretical approach to
quantify the relevant parameters of the localization of a single
target site. Their model describes the overall searching process comprising the
primarily encounter of the enzyme with a DNA domain and the secondary encounter
of the enzyme with the target site. Here we deal with the unvisited case of two
competitive target sites in order to quantitatively analyze the
physical proprieties of the second encounter, i.e. the target site localization
of a restriction enzyme initially bounded to the DNA. Only the
study of such
systems gives access to  the detailed pathway of  secondary encounter
with well defined  initial  conditions. Related experimental studies with two
differentiable target sites located at well--defined positions on
the DNA strand \cite{langowsky83,terry85,stanford00} allows one to handle two descriptive quantities: the
preference and the processivity of the restriction enzymes. The
preference is the ratio of the number of enzymes which react with
one target site, over the number of enzymes which react with the
other target site. The processivity is the fraction of enzymes
which will react successively with the two target sites. To extract
from these experiments physical parameters of the enzyme pathway such as the proportion of time
spent by the enzyme on the DNA, the average number of
dissociation/association events and the average DNA length scanned prior to the target site
localization, it is  necessary to build a reliable physical model
that can mimic the biological situation.

Here, we propose a simple and general stochastic model to describe the
kinetics of target site localization of a restriction enzyme on
DNA, which explicitly combines any 1D motion along the DNA and 3D
excursions in the solution. In the particular case of 1D diffusing motion, our model allows one to recover the
analytic expression for the mean time needed for the enzyme to find a
single target site on DNA given by Berg \& al. \cite{berg81}. This mean time presents an optimum,
corresponding to the quickest finding strategy which can be
discussed in the cases of point--like and extended target sites. The model explicitly
gives the mean number of enzyme visits on the DNA and the proportion of
the DNA visited until the target site is localized. For two target
sites, our model provides theoretical expressions for the
preference and the processivity factors. These expressions involve
two unknown physical parameters: the 1D and 3D residence
frequencies $\lambda$ and $\lambda'$. We show that $\lambda$ is
easily evaluated from the confrontation of the theoretical
preference to experimental data. The second unknown parameter
$\lambda'$, of minor physical relevance, is extracted from the assumption that the searching
strategy is optimal which will be justified. The comparison of the theoretical
processivity factor to experimental data allows us to predict the
value of a dynamic--associated parameter: the probability
that after an excursion the enzyme will associate to the same DNA
substrate it has left, $\pi_r$.

The article is constructed as follows : first we give the general
background of such an approach and we present the hypothesis of
our model. Then we deduce the mean search time from the study of
the density of the first time passage, and for the cases of point--like
and extended target sites we discuss the optimal
strategy in order to find the most quickly as possible the target
site. We give the condition of existence of this optimal strategy as well as its quantitative 
characteristics. We discuss the value of the optimal 1D frequency and
evaluate finite-size effects. Equation \ref{MST} gives the mean target
site localization time for an enzyme which starts from a random
position on the DNA. The complete distribution of the number of visits of the protein on the DNA is explicity determined.
In particular, its mean value is given by Eq. \ref{MNV}. The average number of distinct bp
visited on the DNA 
is given by Eq. \ref{totalscan}. Second, the preference and
the processivity factors of the restriction enzyme for two target
sites, as functions of the distance between the target sites, are
obtained (Eq. \ref{pref} and Eq. \ref{proc}) and compared with
experimental results concerning \textit{Eco}RV \cite{stanford00}.
The comparison gives us the residence time on the DNA per binding
event and other related physical quantities. We then numerically obtain
the mean time needed for the enzyme to go from the first target
site to the second target site (using Eq. \ref{MSTknown}), and the mean
number of visits on the DNA substrate before the two target sites
are cleaved. In conclusion, we discuss the predicted value of $\pi_r$ defined previously.

\bigskip

{\Large \textbf{Model}} 

We present our model in the framework of a generic protein searching
for its target site on the DNA. The case of dimer proteins which
can bind simultaneously to two target sites is not investigated in
order to discard intersegmental transfers. As a first
approximation, the "hopping" translocation process is assumed to
be represented effectively in the 1D diffusion of the protein.
Then, the pathway followed by the protein, considered as a
point-like particle,  is a succession of 1D diffusions along the
DNA strand and 3D excursions in the surrounding solution (Fig.1).
The time spent by the protein on a DNA strand during each binding
event is assumed to follow an exponential law with dissociation
frequency $\lambda$. This law relies on a Markovian
description of the chemical bond which is commonly used. The probability for the protein
  to be still bound to DNA at a random time $t$ (knowing that it is bound at $t=0$) is then
$P(T>t)=\exp(-\lambda t)$, and the probability that the protein
leaves the DNA at a random time $T$ in the interval $[t,t+dt]$ is
$P(t<T<t+dt)=\lambda \exp(-\lambda t)dt$.

The one--dimensional motion on DNA can be modeled from a
continuous Brownian motion  with diffusion coefficient $D$. As it is usually done (see for instance \cite{jeltsch98}), we assume that the
extremities of the DNA chain act on the protein as
\textit{reflecting boundaries}. Thus, a protein when reaching an
extremity during a  binding event is reflected and  continues its
one-dimensional motion. The target site sequence is a specific
sequence of base-pairs (e.g. the restriction enzyme
\textit{Eco}RV, recognizes the sequence $GATATC$ \cite{taylor89}.
The reaction occurs when the reactive domain of the protein
matches the target site sequence. To a first approximation, we
model the target site sequence as being a {\it perfect} reactive
point (Fig.2). The reaction is assumed to be infinitely fast as
soon as the protein meets the target site. Note that in this case
the protein can find the target site only by diffusing along DNA.
The precise mechanism of this elementary act is still subject to
discussion. In particular, the profile of the DNA--protein
interaction potential is unknown, and could be attractive over an
{\it extended} area. It is then reasonable also to treat the case
where the target site is a zone of finite extension $2r$ (Fig. 3).
In that case the target site can then be reached either by
diffusion along the DNA, or by coming directly from a 3D
excursion. This second approach, developed further, gives rise to
strongly different behavior of the search time.

As a first approximation, the excursions are assumed to be
uncorrelated in space. Hence, when dissociating
from DNA, a protein will rebind at a random position. In other
words, the probability to reach a site on DNA after an excursion
is uniformly distributed along the whole DNA molecule. It has been
suggested  \cite{winter81} that for not excessively concentrated long molecules in solution
the DNA
strands form disjoint domains diluted in the medium. A protein
which reaches such a DNA domain will be trapped in it. In this
case excursions might be correlated due to the geometric
configuration of the DNA. As the configuration of a polymer
strand in solution is a random coil, even short three
dimensional excursions can lead to a long effective translocation of
the linear position of the protein on DNA. Consequently, a small
number of long range transitions is sufficient to uncorrelate the
protein position on DNA.

We now introduce three basic quantities used in this work. The
first one, $P_{3D}(t)$, is the probability density that the
protein in the solution at time $t=0$ will bind DNA at time
$t$ at a random position:

\begin{eqnarray}
\displaystyle P_{3D}(t)=\lambda'  \exp(-\lambda' t)
\end{eqnarray}

where the distribution of the time spent during an excursion is
assumed to follow an exponential law with frequency $\lambda'$
corresponding to a mean time spent in the surrounding solution
$\tau'=1/\lambda'$. Accounting rigorously for the entire law is
beyond the scope of this work. Rather we concentrate here on the
characteristic time $\tau'$, which
exists and is finite as soon as the system is confined; and the
exponential tail of the law, which proves to be valid in most plausible
geometries. We will show that this model captures the main relevant characteristics of the
problem.

The second quantity, $P_{1D}(t|x)$, is the conditional probability density that the
protein, being on the DNA at position $x$ and at time $t=0$, will
dissociate at time $t$ without any encounter with the
target site. Assuming that the dissociation rate is independent of the
state of the protein, one has:

\begin{eqnarray}
\displaystyle P_{1D}(t|x)=\lambda \exp(-\lambda t)Q(t|x)
\end{eqnarray}

where $Q(t|x)$ is the conditional probability density that the protein,
starting from the position $x$, does not meet the target site during its
one dimensional diffusion. Introducing $j(t|x)$ as the probability
density of the first passage to the target site position at time $t$
without dissociation,
one gets $Q(t|x)=1-\int_0^tj(t'|x)dt'$.

The last quantity, $\bar{P}_{1D}(t|x)$, is the conditional probability density that
the protein, being on DNA at position $x$ and at time $t=0$, will find
the target site for the first time at time $t$ during its one
dimensional diffusion, without leaving the DNA:

\begin{eqnarray}
\displaystyle \bar{P}_{1D}(t|x)=\exp(-\lambda t)j(t|x).
\end{eqnarray}

Given these quantities, the first passage density of the protein to
the target site can be calculated, first in the case of one target
site, and then we will extend it for two target sites.

\bigskip

\textbf{First passage density}

By calculating the first passage density, we
obtain the mean time needed for the protein to find its specific
target site, as well as all associated moments. We assume that
the  protein  starts at $t=0$ linked to the DNA at position
$x$. We consider a generic event (Fig.2) whose bulk number of
excursions is  $n-1$,  the residence times on
DNA $t_1,\ldots,t_{n}$ and the excursion times
$\tau_1,\ldots,\tau_{n-1}$.  The probability density of such an
event, for which the protein finds the target site  for the first time
t time
$t=\sum_{i=1}^{n}t_i+\sum_{i=1}^{n-1}\tau_i$
is:

\begin{eqnarray}
\displaystyle
P_n(t|x)=\bar{P}_{1D}(t_{n})P_{3D}(\tau_{n-1})P_{1D}(t_n)\ldots
P_{1D}(t_2)P_{3D}(\tau_1)P_{1D}(t_1|x)
\end{eqnarray}

where $P_{1D}(t)$ and $\bar{P}_{1D}(t)$ are averaged over the
initial position of the protein: $P_{1D}(t)=\left<P_{1D}(t|x)
\right>_x$ and ${\bar P}_{1D}(t)=\left<{\bar P}_{1D}(t|x)
\right>_x$.  We denote by $M$ the DNA length on the
``left'' side of the target site  and by $L$ the length on the
``right'' side of the target site. The average of a function $f$ over
the initial position $x$ is given by $\left<f(t|x) \right>_x\equiv
\frac{1}{L+M}\int_{-M}^Lf(t|x){\rm d}x$.

To obtain the density of first passage at the target site,
$F(t|x)$, we sum over all possible numbers of
excursions and we integrate over all intervals of time, ensuring that $t=\sum_i^{n}t_i+\sum_i^{n-1}\tau_i$. The average
over the initial position of the protein, $F(t)=\left<F(t|x)\right>_x$,
can be expressed as:

\begin{eqnarray}
\label{debut}
\displaystyle F(t)=\sum_{n=1}^{\infty}\int_0^{\infty}dt_1\ldots
dt_nd\tau_1\ldots d\tau_{n-1} \delta\left(
\sum_{i=1}^{n}t_i+\sum_{i=1}^{n-1}\tau_i-t) \right) \\
\nonumber \left[ \prod_{i=1}^{n-1} P_{3D}(\tau_i)
\right]\left[\prod_{i=1}^{n-1}P_{1D}(t_i) \right]
\bar{P}_{1D}(t_{n})
\end{eqnarray}

Taking the Laplace transform of $F(t)$,
$\widehat{F}(s)=\int_0^{\infty}dte^{-st}F(t)$,  we obtain:

\begin{eqnarray}
\label{densite} \displaystyle
\widehat{F}(s)=\left<\widehat{j}(\lambda+s|x)\right>_x\;
\left\{1-\frac{1-\left<\widehat{j}(\lambda+s|x)\right>_x}{\left(1+s/
\lambda\right)\left(1+s/ \lambda'\right)}\right\}^{-1}.
\end{eqnarray}

$\widehat{j}(s|x)$ being the Laplace transform of
$j(t|x)$. This expression completely solves our problem {\it for any 1D motion}. We will see in the next section that  the
main quantities of physical interest can be extracted from this
formula.

\bigskip

\textbf{Optimal search strategy}

The relevant quantity to describe the protein/DNA association reaction is the mean
time $\left<\mu\right>$ necessary for the protein to find the
target site (see above). This mean time is obtained from the derivative of the
first passage density by the following relation:

\begin{eqnarray}
\displaystyle
\left<\mu\right>=-\left(\frac{\partial\widehat{F}(s)}{\partial
s}\right)_{s=0}
\end{eqnarray}

which combined with Eq. \ref{densite} gives:

\begin{eqnarray}
\displaystyle
\left<\mu\right>=\frac{1-\left<\widehat{j}(\lambda|x)\right>_{x}}
{\left<\widehat{j}(\lambda|x)\right>_{x}}\left(\frac{1}{\lambda}+\frac{1}{\lambda'}
\right)
\end{eqnarray}
This expression is very general and holds for any $1D$ motion.
Now, we calculate this quantity for a free $1D$ diffusion.
The one--dimensional Laplace transform of the first passage
probability density is well known (see the textbooks
\cite{redner}):
\begin{equation}
{\rm if }\  x>0,
\;\widehat{j}(\lambda|x)=\cosh\left(\sqrt{\frac{\lambda}{D}}x\right)
-\tanh\left(\sqrt{\frac{\lambda}{D}}L\right)\sinh\left(\sqrt{\frac{\lambda}{D}}x\right)
\end{equation}

\begin{equation} \label{jxneg}
{\rm if } \ x<0,
\;\widehat{j}(\lambda|x)=\cosh\left(\sqrt{\frac{\lambda}{D}}x\right)
+\tanh\left(\sqrt{\frac{\lambda}{D}}M\right)\sinh\left(\sqrt{\frac{\lambda}{D}}x\right)
\end{equation}
Averaging over $x$, we finally obtain
\begin{eqnarray} \label{jmoy}
\displaystyle
\left<\widehat{j}(\lambda|x)\right>_{x}=\frac{1}{M+L}\sqrt{\frac{D}{\lambda}}
\left[\tanh\left(\sqrt{\frac{\lambda}{D}}L\right)+\tanh\left(\sqrt{\frac{\lambda}{D}}M\right)\right].
\end{eqnarray}

where $D$ is the one dimensional diffusion coefficient. Then the
mean search time takes the following form:

\begin{eqnarray}
\label{MST} \displaystyle
\left<\mu\right>=\left(\frac{1}{\lambda}+\frac{1}{\lambda'}\right)
\left\{\frac{\sqrt{\frac{\lambda}{D}}(L+M)}{\tanh\left(\sqrt{\frac{\lambda}{D}}L\right)
+\tanh\left(\sqrt{\frac{\lambda}{D}}M\right)}-1\right\}
\end{eqnarray}

Some comments about this expression are in order. First, we recover in a simple and direct way the original result of Berg et al. \cite{berg81}, obtained from a complete description of the 3D motion \cite{berg76,berg77,berg78}.

Second, this
quantity is minimum when the target site is centered (as
expected for symmetry reasons).

Third, as soon as the length of the DNA strand is
large enough (more precisely as soon as
$\sqrt{\frac{\lambda}{D}}L\gg1$ or
$\sqrt{\frac{\lambda}{D}}M\gg1$), $\left<\mu\right>$
grows linearly with the length of the DNA strand. This mirrors the
efficiency of the 1D and 3D combined motion when compared to the
quadratic growth obtained in the case of pure sliding. In
particular, the boundary effects are negligible for this quantity
as soon as the overall length is large enough.

Last, this expression is
valid for a very large class of 3D motions. More precisely, it holds
as soon as the mean first return time $\tau_{3D}$ corresponding to the 
3D motion is finite and independent of the departure and arrival
points. The corresponding expression of the
mean first passage time is obtained by replacing $\lambda'$ by
$1/\tau_{3D}$.

We now come to  an important question, already present in the seminal work of Berg et al.\cite{berg81}  and recently addressed by Slutsky
\& al \cite{mirny04}, which concerns the {\it optimum strategy} for such
a coupled motion. Indeed, it seems reasonable that $\left<\mu\right>$ is large for both
$\lambda$ very large (in the $\lambda$ infinite limit, the protein
 is never on the DNA), and $\lambda$ very small (pure sliding limit).
 It has been  suggested from qualitative
arguments \cite{mirny04} that  the mean search time is minimum when the protein
spends equal times bound to the DNA and freely diffusing in
the bulk. 

Here, we more precisely address this question of minimizing the
mean search time with respect to the 1D frequency $\lambda$. This
is the only specially ``adjustable'' (depending strongly on
the structure of the protein) parameter: $\lambda'$ depends on the
properties of the environment and will not vary significantly
from one protein to another. The 1D diffusion coefficient $D$ is
 a specific quantity, and optimizing the search time
with respect to this parameter is trivial: $D$ should be as large
as possible (note that $D$ and $\lambda$ are assumed to be independent).

The sign of the derivative at $\lambda=0$ of the mean search time
gives the following criterion for having a minimum
\be \label{condex}
\lambda'>15D\;\frac{L^2+M^2-LM}{L^4+M^4+4LM(L^2+M^2)-9M^2L^2} \fin

In fact, it can be shown that this sufficient condition is also
necessary. If this condition is fulfilled, a  careful analysis  of the
implicit equation satisfied by the frequency at the minimum leads
to the following expansion for large $\ell=L+M$

\begin{eqnarray}
\label{lammin}
\displaystyle
\lambda=\lambda'-4\frac{\sqrt{D\lambda'}}{\ell}-\frac{8D}{\ell^2}-\frac{40D^{3/2}}{\sqrt{\
\lambda'}\ell^3}+O\left(\frac{1}{\ell^4}\right).
\end{eqnarray}

Equations \ref{condex} and \ref{lammin} refine the result of Slutsky, which however
holds true in the large $\ell$ limit, or more precisely for
$\sqrt{\frac{\lambda}{D}}\ell\gg1$. For intermediate values of
$\ell$
boundary effects become important and the minimum can be
significantly different.

The $\left<\mu\right>$ value at the minimum is
particularly interesting. We compare it to the case of pure sliding
where $\left<\mu_s\right>=\ell^2/(3D)$: \be
\frac{\left<\mu\right>}{\left<\mu_s\right>}=\frac{6}{\ell}\sqrt{\frac{D}{\lambda}}
\fin The efficiency of the 3D mediated strategy is therefore much more
important when the DNA chain is long. For example, using the $\lambda$
and $D$ values obtained in the section of the results and for a DNA substrate of
length $10^6$ bp, the mean target site localization time when pure sliding is one thousand
fold greater than that predicted by our model.
 
\bigskip

\textbf{Further quantitative features of reactive pathways}

In this paragraph, we compute two quantities which characterize more precisely the nature of the reactive paths. These quantities are of special interest as they could be experimentally measured using single--molecule techniques.
 
The first quantity is the distribution $p(N)$ of the
number of visits on DNA required before reaching the
target site. We recall that in the initial state the protein is
bounded to the DNA, therefore $N\geq1$. The distribution can be obtained by slightly modifying the expression of the first passage density Eq.\ref{debut}:
\be
\begin{array}{lll}
\displaystyle
p(N) & =&\displaystyle \int_0^\infty dt \left<P_N(t|x)\right>_x\\
     & =&\displaystyle \int_0^{\infty}dt\int_0^{\infty}dt_1\ldots
dt_nd\tau_1\ldots d\tau_{n-1} \delta\left(
\sum_{i=1}^{n}t_i+\sum_{i=1}^{n-1}\tau_i-t) \right) \\
  &  & \displaystyle× \left[ \prod_{i=1}^{n-1} P_{3D}(\tau_i)
\right]\left[\prod_{i=1}^{n-1}P_{1D}(t_i) \right]
\bar{P}_{1D}(t_{n})
\end{array}
\fin
Finally, this distribution  happens to be a geometric law with parameter
$\left<\widehat{j}(\lambda|x)\right>_{x}$: 
\be
p(N)=\left<\widehat{j}(\lambda|x)\right>_{x}\left(1-\left<\widehat{j}(\lambda|x)\right>_{x}\right)^{N-1}
\fin 
This demonstrates that the mean number of visits
before reaching the target site is: \be \label{MNV}\left<N\right>=\frac{1}{\left<\widehat{j}(\lambda|x)\right>_{x}} =\frac{\sqrt{\frac{\lambda}{D}}(L+M)}{\tanh\left(\sqrt{\frac{\lambda}{D}}L\right)
+\tanh\left(\sqrt{\frac{\lambda}{D}}M\right)}\fin 
The following form holds:
\be \label{MNVbis}
\left<\mu\right>=\left(\left<N\right>-1\right)\left(\frac{1}{\lambda}+\frac{1}{\lambda'}\right)\fin
Note that the large $N$ limit is transparent ($\left<\mu\right>$ is a succession of approximately $N$ 1D excursions of average duration $1/\lambda$ and  $N$ 3D excursions of average duration $1/\lambda'x$).

The second interesting quantity is the average number of distinct base pairs visited before the protein reaches its target site. In our continuous description, this corresponds to the average span $\left<S\right>$ of the 1D motion.  For sake of simplicity, the target is here assumed to be centered on the DNA strand of half length $L$. The average span can be expressed as the integral over the position $x$ on the DNA of the probability that $x$ has been visited before reaction. One then obtains:
\be
 \left<S\right>=\int_{-L}^Ldx\int_{0}^\infty dt F_{\bar 0}(x,t)=\int_{-L}^Ldx {\hat F}_{\bar 0}(x,s=0)
\fin
where $ F_{\bar 0}(x,t)$ is the first passage density at $x$ with adsorbing conditions at $x=0$, whose Laplace transform will be explicitly computed in the next section in the context of competitive targets. Anticipating formula Eq.\ref{densite2}, the span finally reads:
\be \label{totalscan}
\left<S\right>=2\int_{0}^{L}dx \left\{1+\frac{\cosh\left(\sqrt{\frac{\lambda}{D}}(L-x)\right)\sinh\left(\sqrt{\frac{\lambda}{D}}(L+x/2)\right)}{\cosh\left(\sqrt{\frac{\lambda}{D}}L\right)\sinh\left(\sqrt{\frac{\lambda}{D}}(L-x/2)\right)}\right\}^{-1}
\fin
Apparently,  this integral form can not be substantially simplified,
but its overall behaviour, and in particular the $\lambda$ dependence, is
easily cleared up. The span appears to grow monotonously from
$\frac{3}{4}L$ at $\lambda=0$ to $L$ for $\lambda\to\infty$. This monotonicity, as
opposed to the existence of a minimum for the mean search time,  is a
striking feature of this quantity, plotted in figure 6.

\bigskip

\textbf{Extended target site}

As mentioned above, the model of a point--like target site disregards the
possibility of the protein reaching the target site directly from a 3D
excursion. For this reason, we have to study 
the case where the target site is an area of extension $r$. We will now show that this new feature significantly changes the behaviour of the searching time. The
reaction is still assumed to be infinitely fast; it occurs either
when the protein reaches the boundary of the reaction area during
a sliding round, or when the protein comes on the
reaction area directly after a 3D excursion. Following the scheme already
developed to derive the density of the first passage time
(\ref{densite}), one obtains:

\begin{eqnarray}
\label{densiteetendue} \displaystyle
\widehat{F}(s)=\left(\left<\widehat{j}(\lambda+s|x)\right>_*
+\frac{2r}{L+M}\right)\;\left\{1-\frac{1-\frac{2r}{L+M}-\left<\widehat{j}(\lambda+s|x)\right>_*}{\left(1+s/
\lambda\right)\left(1+s/ \lambda'\right)}\right\}^{-1}.
\end{eqnarray}
where $\left<f\right>_*=\frac{1}{L+M}(\int_{-M}^{-r}f
dx+\int_{r}^{L}f dx)$. The average search time then reads (we
only give the case $L=M$ for  sake of simplicity): \be
\left<\mu(r)\right>=\left(\frac{1}{\lambda}+\frac{1}{\lambda'}\right)\frac{(\ell-r)\sqrt{\frac{\lambda}{D}}
-\tanh((\ell-r)\sqrt{\frac{\lambda}{D}})}{r\sqrt{\frac{\lambda}{D}}+\tanh((\ell-r)\sqrt{\frac{\lambda}{D}})}
\fin For $\ell$ large enough, the minimum is obtained for \be
\lambda_{min}\simeq \frac{ \left(\lambda'\,r+\sqrt
{{\lambda'}^{2}{r}^{2}+D\,\lambda'} \right)^{2}}{D} \fin It is
remarkable that  the scaling $\lambda_{min}\approx\lambda'$ holds true
only for $\lambda'\ll D/r^2$. For larger frequencies $\lambda'$,
we have $\lambda_{min}\approx 4\lambda'^2r^2/D$. The value of the
search time at the minimum $\left<\mu(r)\right>_{min}$ is
modified. For $r$ small we get: \be
\left<\mu(r)\right>_{min}=\frac{2\ell}{\sqrt{\lambda'D}}-\frac{2\ell
r}{D}+{\cal O}(r^2) \fin whereas for larger $r$ the expansion
reads: \be \left<\mu(r)\right>_{min}=\frac{\ell}{\lambda'
r}-\frac{D\ell}{4\lambda'^2r^3}+{\cal O}(1/r^5) \fin

We now consider the case of two target sites in order to compare the model
to experimental results.

\bigskip

{\large \textbf{Case of two competitive target sites}}

The biological system \cite{stanford00} consists in integrating
two target sites for the restriction enzyme \textit{Eco}RV on a
690 bp linear DNA substrate. The position along a DNA strand of
the first target site, which will be called target $1$, is fixed
and equals 120 bp. The second target site, which will be called
target $2$, has been placed at 54 bp, 200 bp, and 387 bp from the
first target site. Thus, three substrates (Fig.5) were used to
analyze the kinetics of DNA cleavage. Each assay was carried out
at a very low concentration of enzyme with regard to the
concentration of DNA . For higher concentration of enzyme, the
probability for two --or more-- molecules acting on a same DNA
strand would be not negligible. The cleavage of DNA produces
different lengths of DNA. An enzyme can cut target $1$, target
$2$, or both, resulting in 5 lengths of fragments. The authors
observed the initial formation of four of these: $A$, $BC$, $C$
and $AB$ types.

The advantage of this construction is that the first cleavage process gives a starting point to elucidate how
\textit{Eco}RV will cleave the second target site. In contrast, when using
constructions with one target site, the primary pathway of the
enzyme to reach the DNA domain can dominate the kinetics of the
search process. For example, in highly diluted DNA solutions, the DNA domains are separated by long distances and
then the mean time spent by the enzyme in reaching a DNA domain will
contribute in a nonegligible manner to the total mean time needed
to find the target site. Moreover, our theoretical model supposes that
the enzyme starts on the DNA and therefore does not comprise the
primary encounter. This assumption agrees with the case of
experimental substrates with two target sites .

\bigskip

\textbf{Conditional search time density}

In order to get a better understanding of this process we first study
analytically the distribution of the search time $t$ of one target,
for instance 2, knowing that no reaction occurred at target 1. We
denote by  $F_{\bar 1}(2,t)$ this  conditional search time density
averaged over the initial condition. We make use of the general method
developed in the first section to derive this quantity. Indeed, this
problem involves a combination of 3D excursions and 1D motions, its
peculiarity being that the 1D motion is a constrained diffusion, as
reaction with target 1 is excluded. It suffices then to rewrite formula Eq.\ref{densite} as follows:

\begin{eqnarray}
\label{densite2} \displaystyle
\widehat{F}_{\bar 1}(2,s)=\left<\widehat{j}_{\bar 1}(\lambda+s|2,x)\right>_x\;
\left\{1-\frac{1-\left<\widehat{j}_{\bar 1}(\lambda+s|2,x)\right>_x-\left<\widehat{j}_{\bar 2}(\lambda+s|1,x)\right>_x}{\left(1+s/
\lambda\right)\left(1+s/ \lambda'\right)}\right\}^{-1}
\end{eqnarray}

The first factor $\left<\widehat{j}_{\bar 1}(s|2,x)\right>_x$ is  the Laplace transform of the first passage density at 2 avoiding 1 for a standard 1D diffusion, and corresponds to the last excursion before finding the target 2. In turn, the term proportional to $\left(1-\left<\widehat{j}_{\bar 1}(\lambda+s|2,x)\right>_x-\left<\widehat{j}_{\bar 2}(\lambda+s|1,x)\right>_x\right)/s$ is the Laplace transform of the survival probability density, and comes from the succession of non reactive excursions on DNA. Theses  quantities are obtained by standard methods, considering successively the initial condition on fragment $A$ (with mixed boundary conditions), $B$ (with absorbing boundary conditions), and $C$ (with mixed boundary conditions). This finally yields to 

\begin{equation}
\displaystyle
\left<\widehat{j}_{\bar 1}(\lambda|2,x)\right>_x=\frac{1}{\ell}\sqrt{\frac{D}{\lambda}}\left\{\tanh\left(\sqrt{\frac{\lambda}{D}}c\right)
+\frac{\cosh\left(\sqrt{\frac{\lambda}{D}}b\right)-  1}{\sinh\left(\sqrt{\frac{\lambda}{D}}b\right)}\right\}
\end{equation}
and 
\begin{equation}
\displaystyle
\left<\widehat{j}_{\bar 2}(\lambda|1,x)\right>_x=\frac{1}{\ell}\sqrt{\frac{D}{\lambda}}\left\{\tanh\left(\sqrt{\frac{\lambda}{D}}a\right)
+\frac{\cosh\left(\sqrt{\frac{\lambda}{D}}b\right)-  1}{\sinh\left(\sqrt{\frac{\lambda}{D}}b\right)}\right\}
\end{equation}
where $a,b,c$ denote the length of fragments $A,B,C$ respectively. This set of equations fully describes the problem, and will be used in next section to analyze experimental data. In particular the mean conditional search time could be deduced straightforwardly from Eq. \ref{densite2}; its explicit form is not given here for sake of simplicity.

\bigskip

\textbf{Preference and processivity}

In order to get quantitative measurements of the pathway of the
enzyme, the authors of \cite{stanford00} introduced two concepts: preference and
processivity. The value of the preference $P$ quantifies the
preferential use of the target $2$ by \textit{Eco}RV. The $P$ value is experimentally obtained
by taking the ratio of the initial formation rate $\nu_{AB}$ of
$AB$ substrates (resulting from cleavage at the target site $2$),
over the initial formation rate $\nu_{BC}$ of $BC$ substrates (resulting
from cleavage at the target site $1$).

\begin{eqnarray}
\displaystyle P=\frac{\nu_{AB}}{\nu_{BC}}
\end{eqnarray}

The processivity quantifies the fraction of the cleaved DNA
that is cleaved first at one target site then cleaved at the
second target site during the encounter of the DNA substrate with an
enzyme. The processivity of the restriction enzyme on the target $2$ to
the target $1$, can be deduced from experimental
data by  introducing the processivity factor:
$f_{p21}=(\nu_C-\nu_{AB})/(\nu_C+\nu_{AB})$. One can define in the same manner a symmetric quantity, which is the
processivity factor of the reaction with the target $1$ and then
target $2$: $f_{p12}=(\nu_A-\nu_{BC})/(\nu_A+\nu_{BC})$
 and then the total processivity factor which represent the
fraction of both processive actions:

\begin{eqnarray}
\displaystyle
f_p=\frac{\nu_A+\nu_C-\nu_{AB}-\nu_{BC}}{\nu_A+\nu_C+\nu_{AB}+\nu_{BC}}.
\end{eqnarray}

The next sections deal with these two
quantities obtained from our model by considering the enzyme-to-target(s)
association rate, namely $\nu_1$, $\nu_2$, $\nu_{21}$, and
$\nu_{12}$ which are defined by the following elementary
reactions, instead of substrate rate production:

 \be
\begin{array}{lll}
DNA\longrightarrow & A+BC & {\rm with\ rate\ }\nu_1\\
DNA\longrightarrow & AB+C & {\rm with\ rate\ }\nu_2\\
DNA\longrightarrow & A+BC \longrightarrow A+B+C & {\rm with\ rate\ }\nu_{21}\\
DNA\longrightarrow & AB+C \longrightarrow A+B+C & {\rm with\ rate\
}\nu_{12}

\end{array}
\fin

We assume that a restriction enzyme hits a DNA molecule at site $x$ with
homogeneous probability per unit time $\kappa dx/(L+M)$. The
enzyme concentration is chosen sufficiently small so that
multiple encounter events are negligible. Consequently, a fragment
$BC$ (or $AB$) can be cut into $B$ and $C$ (or $A$ and $B$) only
if the enzyme which cleaves the DNA molecule to give $BC$ (or $AB$) remains on this fragment (the probability of this
event, depending in detail on the chemical mechanism, will be
denoted $p_{init}$) and then finds the site $2$ (or $1$). The
reaction rates is then:
\be \nu_1=\kappa\int_{-\infty}^t
dt'\int_{DNA} \frac{dx}{L+M} F_{\bar 2}(1,x, t-t')
=\kappa\left<{\hat F}_{\bar 2}(1,x,s=0)\right>_x \fin and
\be\begin{array}{lll}\disp \nu_{12}& = & \disp \kappa
p_{init}\int_{-\infty}^t dt' F(1,2, t-t')\int_{-\infty}^{t'}
dt''\int_{DNA} dx
 F_{\bar 1}(2,x, t'-t'')\\
    & = & \kappa p_{init}{\hat F}(1,2,s=0)\left< {\hat F}_{\bar 1}(2,x, s=0)\right>_x
\end{array}
\fin where the quantity $F_{\bar z}(y,x, t)$ is the first passage
density at point $y$ at time $t$ starting from $x$ and avoiding
$z$. This quantity  is accessible analytically using Eq.\ref{densite2}. The quantity $F(y,x,t)$ is
the first passage density at point $y$ at time $t$ starting from
$x$. The two other rates $\nu_{2}$ and $\nu_{21}$ are
straightforwardly obtained by permutation of symbols 1 and 2. One
is now able to derive the processivity and preference factors.

\bigskip

{\large \textbf{Results}}

We recall that the lengths of fragments $A$, $B$ and $C$ are denoted by the
lower--case letters $a$, $b$ and $c$ respectively. First, we
evaluate the 1D frequency $\lambda$ from the comparison of the
theoretical preference to experimental data. Then, using the value
of $\lambda'$ which satisfies the optimal searching time (this assumption is justified below), we
deduce several quantities related to the enzyme pathway which
links the first target site to the second one. Last, by comparing
the analytical expression of the processivity factor to
experimental data, we introduce a dynamic--associated
parameter: the probability that after an
excursion the enzyme will associate to the same DNA substrate it
has left, $\pi_r$.

\textbf{Preference}

The {\it preference} for the target site 1 over site 2 is given by
\be
P=\frac{E_2}{E_1}=\frac{\nu_{AB}}{\nu_{BC}}=\frac{\nu_2-\nu_{12}}{\nu_1-\nu_{21}}=
\frac{\left< {\hat F}_{\bar 1}(2,x, s=0)\right>_x}{\left< {\hat
F}_{\bar 2}(1,x, s=0)\right>_x} \fin where $\nu_x=dx/dt$ is the
rate for forming the specie $x$, which can be measured
experimentally. Explicitly: \be \label{pref}
P=\frac{\tanh(\sqrt{\frac{\lambda}{D}}c)+
(\cosh(\sqrt{\frac{\lambda}{D}}b)-1)/\sinh(\sqrt{\frac{\lambda}{D}}b)}{\tanh(\sqrt{\frac{\lambda}{D}}a)+
(\cosh(\sqrt{\frac{\lambda}{D}}b)-1)/\sinh(\sqrt{\frac{\lambda}{D}}b)}
\fin

This form which expresses the preference as function of $b$, and reveals in
particular that the preferred target site
 is the closest to the middle of the molecule, well fits the experimental
data (Fig.7)
 and allows one to determine the only free parameter $\sqrt{{\lambda}/{D}}$. The best fit is
obtained for: $\sqrt{{\lambda}/{D}}=8.7.10^{-2}$ $bp^{-1}$. For a representative
fast one-dimensional diffusion coefficient $D=5.10^5$
$bp^2/s$ \cite{erskine97}, the 1D frequency is $\lambda=37.5$ $s^{-1}$. Then the average time spent
by the restriction enzyme on DNA per visit equals $0.027$ s and the average
distance scanned per visit ($\sqrt{16D/\pi \lambda}$) is 260
bp. Using Eq. \ref{totalscan}, we obtain a
representative average number of distinct sites
visited on the DNA during the searching process: $<S>\simeq320bp$.

\textbf{Enzyme pathway}

A further analysis requires to know the value of the parameter $\lambda'$, which depends strongly on experimental conditions, such as DNA concentration. It could be obtained experimentally as the protein/DNA association rate, and we here choose a typical value corresponding to the optimal search strategy, i.e. $\lambda=\lambda'$.  This assumption is supported by the fact that the target site localization is several order of magnitude
faster than the diffusion limit. Using the same calculation as from Eq. \ref{debut} to
Eq. \ref{MST} without averaging on the initial position of the enzyme,
we obtain the mean time needed by the restriction enzyme to go
from the target $1$ to the target $2$:

\be \label{MSTknown}
\left<\mu\right>=\left(1-\frac{1}{\cosh(\sqrt{\frac{\lambda}{D}}b)}\right)\left(\frac{(b+c)\left(\frac{1}{\lambda}+\frac{1}{\lambda'}\right)}{\tanh(\sqrt{\frac{\lambda}{D}}b)+\tanh(\sqrt{\frac{\lambda}{D}}c)}-\frac{1}{\lambda'}\right)
\fin

Then the average search time of the target $2$
for a reactive pathway of an enzyme starting from the target
$1$, with inter--site space of $54$ bp, is by using the formula
\ref{MSTknown}: $\left<\mu\right>\simeq0.016s$. The average number of
DNA visits before the processive cleaving is, using the
formula \ref{MNVbis}, $N\simeq1.3$. . The same quantities for the other
inter-target site distances, namely $200$ bp and $387$ bp, are
respectively: $\left<\mu\right>\simeq0.072s$, $N\simeq2.4$; and
$\left<\mu\right>\simeq0.10s$, $N\simeq2.9$. 

\textbf{Processivity}

Using the previous results, the processivity factor takes the
following form: \be f_p=\frac{\nu_{12}+ \nu_{21}}{ \nu_{1}+
\nu_{2}}=  p_{init}{\hat F}(1,2,s=0) \fin Here we have to refine
the derivation of ${\hat F}(1,2,s=0)$, i.e. the probability to
ever reach 1 starting from 2. The crucial point is about the
dilution approximation, hence we treat the case of one single
enzyme. We take into account the fact that during each
 $3D$ excursion the protein can escape, therefore being definitely lost. We introduce
 by
 $\pi_r$ the probability of return after a 3D excursion. 
Rigorously this quantity depends on physical parameters such as the
DNA length and the typical size of its  attractive domain. As the
lengths of DNA substrates are constant in the experiments of Stanford et
al. \cite{stanford00} for which $b+c=570bp$, we consider a constant $\pi_r$. We
finally obtain:
\be \label{proc}
f_p=p_{init} \left( \widehat{j}(\lambda|2,1)+\frac{\pi_r\left<\widehat{j}(\lambda|x)\right>_x(1-\widehat{j}(\lambda|2,1))}{1-\pi_r+\pi_r\left<\widehat{j}(\lambda|x)\right>_x}  \right) 
\fin
Where $\left<\widehat{j}(\lambda|x)\right>_x$ is given by the
Eq. \ref{jmoy} with $L=c$ and $M=b$, and where $\widehat{j}(\lambda|2,1)$ is the Laplace transform of the first
passage density at 2, starting from 1 which is given by
Eq. \ref{jxneg} with $x=M=b$:
\be 
\widehat{j}(\lambda|2,1)=\cosh(\sqrt{\frac{\lambda}{D}}b)
\fin

Using the value of $\lambda$ obtained previously, there are 2 unknown
parameters:
 $p_{init}$ and $\pi_r$. They can be determined from the
 experimental data (Fig.8); the best fit is obtained for $p_{init}=0.5$ and
 $\pi_r=0.85$. However, these values can be not very accurate as it is used to
be the case when estimating two parameters by fitting experimental
data with theoretical results.

 We will discuss some
 possible hypotheses arising from the two last fitted parameters in the following conclusion.

\bigskip

{\large \textbf{Conclusion}}

So far, experimental investigations have allowed one to
discriminate between two translocation processes, pure sliding or
pure jumping. To obtain quantitative measurements for such a
compound translocation process, it is necessary to build a physical
reliable model, as
Berg \& al. \cite{berg81} did for a single target site. The model presented here permits
us to obtain numerous quantities determining the pathway followed by a restriction enzyme in
finding one target site or two competitive target sites on DNA, by a series of 1D
diffusion periods (sliding) followed by 3D excursions (jumping).
The corresponding mean search time shows that such a two-step process
is faster than pure sliding or pure 3D diffusion. The existence and
the optimization
of such a search time is discussed. The length dependence of the
optimum was obtained.

Using the preference data from assays on
\textit{Eco}RV \cite{stanford00}, we quantify
the parameter characterizing the pathway of \textit{Eco}RV, namely the 1D
residence frequency $\lambda$. Other quantities were
extracted from this parameter: the mean distance scanned by the
restriction enzyme during one binding event (260 bp), the distribution
of the number of visits on DNA prior to cleaving the target site and the
average number of distinct DNA sites visited. It should be noticed that
the small value of the mean distance scanned might be due to the
assumption of a perfect reactive target site which leads to an over--estimated $\lambda$. In fact, an imperfect
reactive target site would decrease the preference. Using the data on processivity
for \textit{Eco}RV, we introduce two secondary parameters
characterizing the detailed pathways of the
restriction enzyme after DNA cleavage. These parameters
come into play when more than one target site is present on the DNA. The first parameter is the probability for the enzyme
 to stay (after cleavage with a target site) on the DNA strand which
harbors the second target site. It was assumed that this probability
equals $1/2$ as the DNA sequences which border the target site are
almost symmetric. Our best fit suggest that the probability is
fairly 0.5, justifying the common assumption. The
second parameter $\pi_r$ is the probability for the
enzyme to rebind on the cleaved DNA strand it
had left during an excursion. Because of the short length of DNA substrates, it is
assumed that the enzyme is "lost" after the dissociation from the
DNA. This means that the enzyme rebinds unvisited DNA substrates
after each 3D excursion. Therefore, this probability had been previously assumed to
be negligible. Our model reveals that this probability is
high (0.85) which shows that the enzyme
frequently rebinds to the same DNA substrate. The high value of $\pi_r$
may be explained by the fact that the fragment length $\ell$ (which is
here $b+c=570 bp$) is significantly larger than the persistence
length (150 bp). The configuration of the DNA is therefore close to a
globule, in which the protein can be trapped and hence escape with a rather
low probability.  However, $\pi_r$ may be overestimated because of our
assumption of neglecting the correlations between the starting and
finishing points of the 3D excursions. Indeed, these correlations
would result (for small values of the inter-target distance $b$) in
increasing the processivity factor, and therefore lowering $\pi_r$. Note
that an imperfect reaction would lower the processivity, as in this
case the enzyme can pass trough the target site without react, therefore
increasing the probability of a definitive departure from the DNA strand.

The present model classifies the stochastic
pathway followed by a restriction enzyme searching for its target site,
by quantifying the dynamical parameters. Our work is in the framework of stochastic dynamics which
dictates the biological processes occurring in the highly
structured and crowded medium of in-vivo systems. Moreover, this
model can be helpful for generic situations where a
protein has to find a target site on a DNA substrate, e.g. the
numerous transcription factors needed to trigger the gene activation.

\bigskip

\small{We are grateful to M. Barbi, G. Oshanin, and J.M. Victor
(LPTL) for useful discussions. We are also grateful to J. Coppey and
M. Jardat
for specific comments on the manuscript. The numerous pertinent comments, criticisms and suggestions 
given
by one referee were  deeply appreciated.}

\newpage


\begin{thebibliography}{99}

\bibitem[Berg, et al., 1976]{berg76} Berg, O.G., and Blomberg,
C. 1976. Association kinetics with coupled diffusional flows. Special
application to the lac repressor--operator system. Biophys. Chem.,
\textbf{4}, 367-381.

\bibitem[Berg, et al., 1977]{berg77} Berg, O.G., and Blomberg,
C. 1977. Association kinetics with coupled diffusion. An extension to
coiled--chain macromolecules applied to the lac repressor--operator system. Biophys. Chem.,
\textbf{7}, 33-39. 

\bibitem[Berg, et al., 1978]{berg78} Berg, O.G., and Blomberg,
C. 1978. Association kinetics with coupled diffusion
III. Ionic--strength dependence of the lac repressor--operator
association. Biophys. Chem.,
\textbf{8}, 271-280.

\bibitem[Berg, et al., 1981]{berg81} Berg, O.G., Winter, R.B. and von
Hippel, P.H. 1981. Diffusion-driven mechanisms of protein translocation on nucleic acids. 1. Models and theory. Biochemistry, \textbf{20}, 6929-6948.

\bibitem[Erskine, et al., 1997]{erskine97} Erskine, SG, Baldwin, GS,
Halford, SE. 1997.
Rapid-reaction analysis of plasmid DNA cleavage by the EcoRV
restriction endonuclease. Biochemistry, \textbf{36}(24), 7567-76.

\bibitem[Halford, et al., 2002]{halford02}  Halford, SE, Szczelkun,
MD. 2002. How to get from A to B: strategies for analysing protein motion on DNA. Eur Biophys J.;\textbf{31}(4):257-67. Review.

\bibitem[Jeltsch, et al., 1998]{jeltsch98} Jeltsch, A. and Pingoud,
A. 1998. Kinetic characterisation of linear diffusion of the restriction endonuclease EcoRV on DNA. Biochemistry, \textbf{37}, 2160-2169.

\bibitem[Langowski, et al., 1983]{langowsky83}  Langowski, J, Alves,
J, Pingoud, A, and Maass, G. 1983. Free in PMC Does the specific recognition of DNA by the restriction endonuclease EcoRI involve a linear diffusion step? Investigation of the processivity of the EcoRI endonuclease. Nucleic Acids Res. \textbf{11}(2):501-13.

\bibitem[Lohman, 1986]{lohman96} Lohman, T.M. 1996. Kinetics of protein-nucleic acid interactions: use of salt effects to probe mechanisms of interactions. CRC Crit. Rev. Biochem., \textbf{19}, 191-245.

\bibitem[Milsom, et al., 2001]{milsom01} Milsom, SE, Halford, SE,
Embleton, ML, Szczelkun, MD. 2001. Analysis of DNA looping interactions by type II restriction enzymes that require two copies of their recognition sites. J Mol Biol \textbf{31}:517-528

\bibitem[Misteli, 2001]{mistelli01} Misteli, T. 2001. Protein dynamics: implications for nuclear architecture and gene expression. Science \textbf{291}:843-847

\bibitem[Redner, 2001]{redner} Redner, S. 2001. A Guide to First-Passage
Processes. Cambridge University Press; pp: 307, ISBN 0-521-65248-0.

\bibitem[Richter, et al., 1974]{richter74} Richter, P.H. and Eigen,
M. 1974. Diffusion controlled reaction rates in spheroidal
geometry. Application to repressor--operator association and membrane
bound enzymes. Biophys. Chem.,
\textbf{2}, 255-263.

\bibitem[Riggs, et al., 1970]{riggs70} Riggs, A.D., Bourgeois, S. and
Cohn, M. 1970. The lac repressor--operator interaction. III. Kinetic
studies. J. Mol. Biol., \textbf{53}, 401-417.

\bibitem[Shimamoto, 1999]{shimamoto99} Shimamoto, N. 1999. One-dimensional diffusion of proteins along DNA. J. Biol. Chem., \textbf{274}, 15293-15296.

\bibitem[Slutsky, et al., 2004]{mirny04}  Slutsky, M., and Mirny, L.A.
2004. How does a protein find its site on DNA? q-bio.BM/0402005 condmat.

\bibitem[Stanford, et al., 2000]{stanford00}  Stanford, NP, Szczelkun,
MD, Marko, JF, and Halford, SE. 2000.One- and three-dimensional pathways for proteins to reach specific DNA sites. EMBO J., \textbf{19}:6546-57.

\bibitem[Szczelkun, et al., 1996]{szczelkun96} Szczelkun, MD, and Halford,
SE. 1996. Recombination by resolvase to analyse DNA communications by the
SfiI restriction endonuclease. EMBO J \textbf{15}:1460-1469

\bibitem[Taylor, et al., 1989]{taylor89} Taylor, J.D. and Halford,
S.E. 1989. Discrimination between DNA sequences by the EcoRV restriction endonuclease. Biochemistry, \textbf{28}, 6198-6207.

\bibitem[Terry, et al., 1985]{terry85} Terry, B.J., Jack, W.E. and
Modrich, P. 1985. Facilitated diffusion during catalysis by EcoRI endonuclease. J. Biol. Chem., \textbf{260}, 13130-13137.

\bibitem[Von Hippel, et al., 1989]{vonhippel89} Von Hippel, P.H. and
Berg, O.G. 1989. Facilitated target location in biological systems. J. Biol. Chem., \textbf{264}, 675-678.

\bibitem[Wenner, et al., 1999]{wenner99} Wenner, J.R., and Bloomfield,
V.A. 1999. Crowding
Effects on \textit{Eco}RV Kinetics and Binding. Biophys. J. 77: 3234-3241.

\bibitem[Winter, et al., 1981]{winter81} Winter, R.B., Berg, O.G. and
von Hippel, P.H. 1981. Diffusion-driven mechanisms of protein translocation on nucleic acids. 3. The Escherichia coli lac repressor--operator interaction: kinetic measurements and conclusions. Biochemistry, \textbf{20}, 6961-6977.

\end{thebibliography}

\newpage

{\large \textbf{Figure legend}}

\textbf{Figure 1.} A representative path of the restriction enzyme
which reaches the target site. Excursions in the solution are
represented by dashed lines, one-dimensional diffusion by
continuous lines. The filled square is the target site.

\textbf{Figure 2.} Representative view of the model.
Here the the protein executes three excursions before finding the
target site.

\textbf{Figure 3.} Extended target site.

\textbf{Figure 4.} Schematic representation of the
three substrates of length 690 bp. The position of the second
target site relative to the first target equals 54 bp, 200 bp and 387
bp, respectively.

\textbf{Figure 5.} The mean search time plotted
against the one-dimensional residence frequency $\lambda$. The length
of DNA is 5000 bp, the three-dimensional residence
frequency is $10s^{-1}$ and the $1D$ diffusion coefficient
is $5.10^5bp^2/s$.

\textbf{Figure 6.} The average number of distinct
DNA sites visited by the enzyme against the one-dimensional residence frequency $\lambda$. The half--length of DNA is 100bp which
allows one to also read this number as a percentage.

\textbf{Figure 7.} The preference of the protein for
the target site $2$ over the target site $1$. The solid line represents the
fitted solution which gives $\sqrt{{\lambda}/{D}}=8.7.10^{-2}bp^{-1}$. The two dashed
lines correspond to the limit cases when there is no sliding
(straight line, $\lambda=\infty$) and when there is only sliding
(upper line, $\lambda=0$). The other parameters were drawn from experimental data ($\ell=690bp$).

\textbf{Figure 8.} The processive action of the
restriction enzyme. Dashed lines represent two fitted solutions of the model
of Stanford \cite{stanford00} with pure sliding.
The two solid lines represent the solutions of our model for
$\sqrt{{\lambda}/{D}}=8.7.10^{-2}bp^{-1}$ and $p_{init}=0.5$:
one for $\pi_r=0$, and the other one which passes near experimental points for $\pi_r=0.85$.

\newpage

{\large \textbf{Figures}}

\begin{figure}[ht]
\begin{center}
\includegraphics*[scale=0.5]{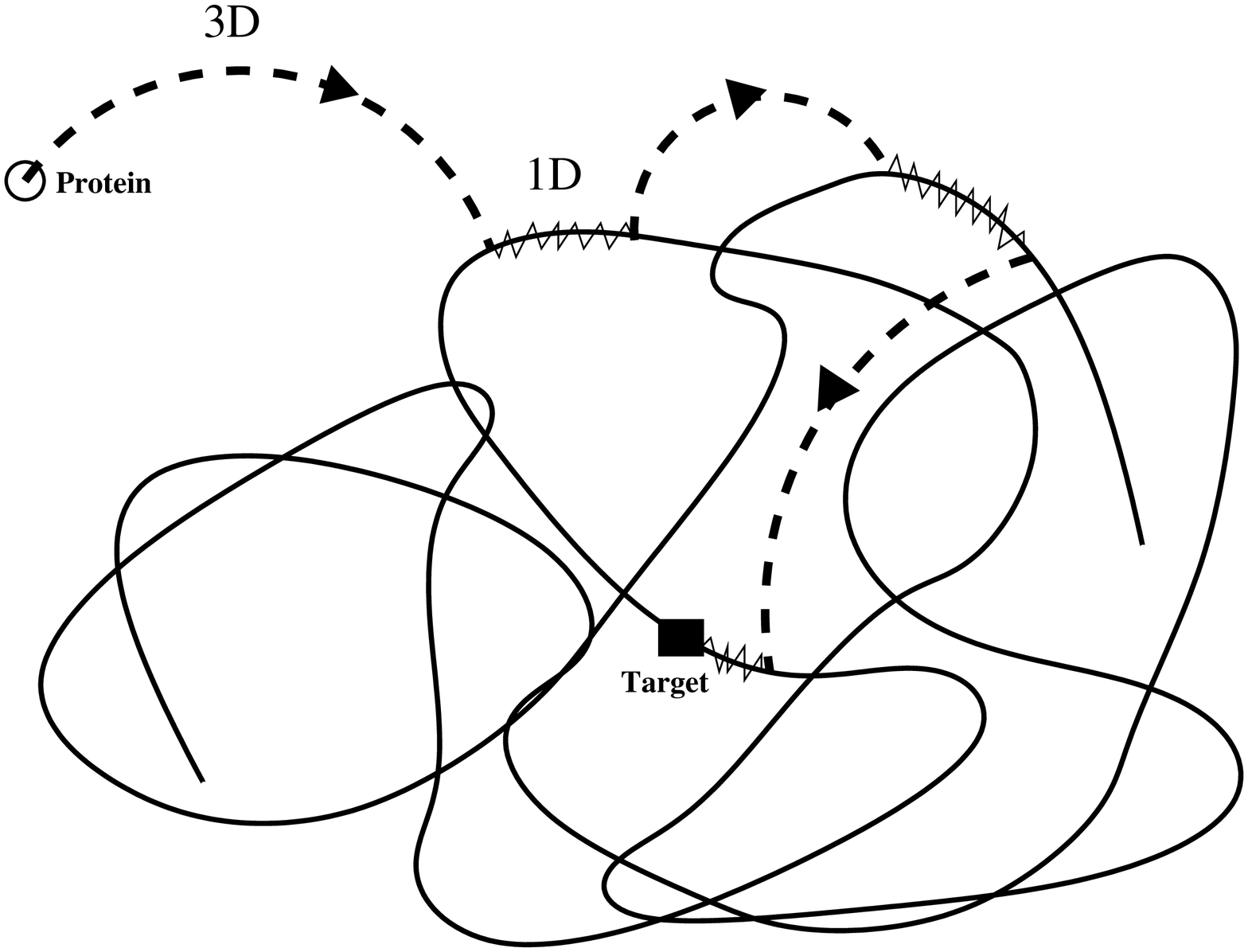}
\caption{\label{sch1}}
\end{center}
\end{figure}

\begin{figure}[ht]
\begin{center}
\includegraphics*[scale=0.6]{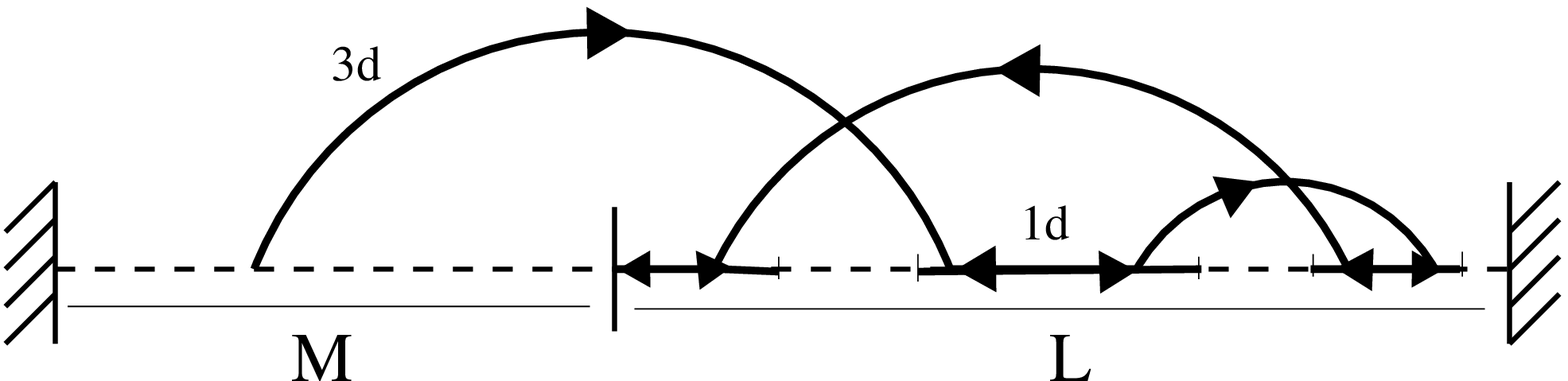}
\caption{\label{sch2}}
\end{center}
\end{figure}

\begin{figure}[ht]
\begin{center}
\includegraphics*[scale=0.6]{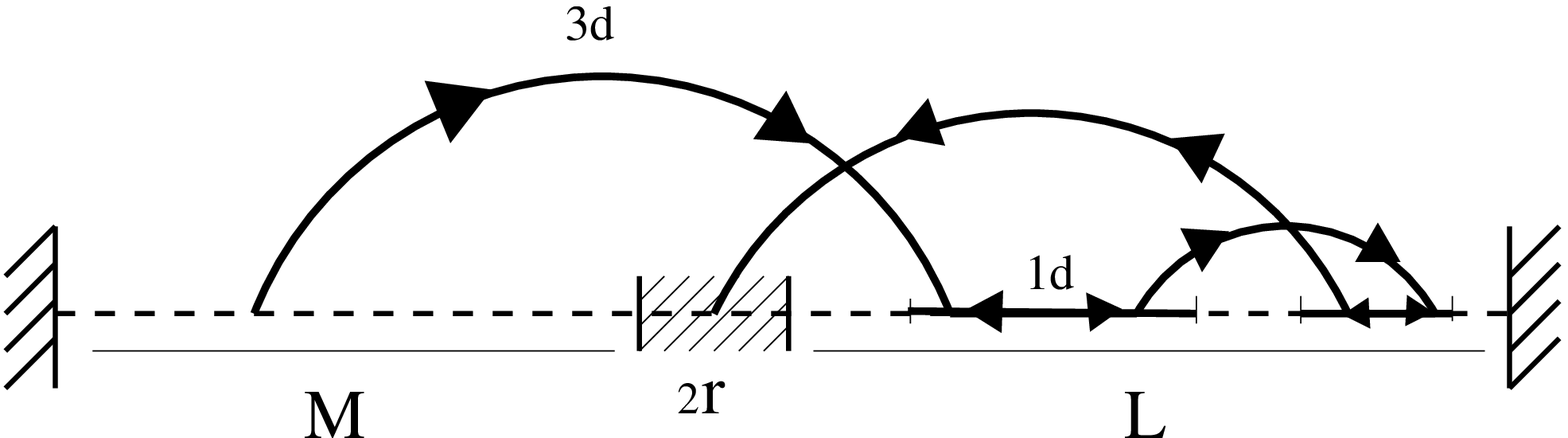}
\caption{\label{sch3}}
\end{center}
\end{figure}

\begin{figure}[ht]
\begin{center}
\includegraphics*[scale=0.6]{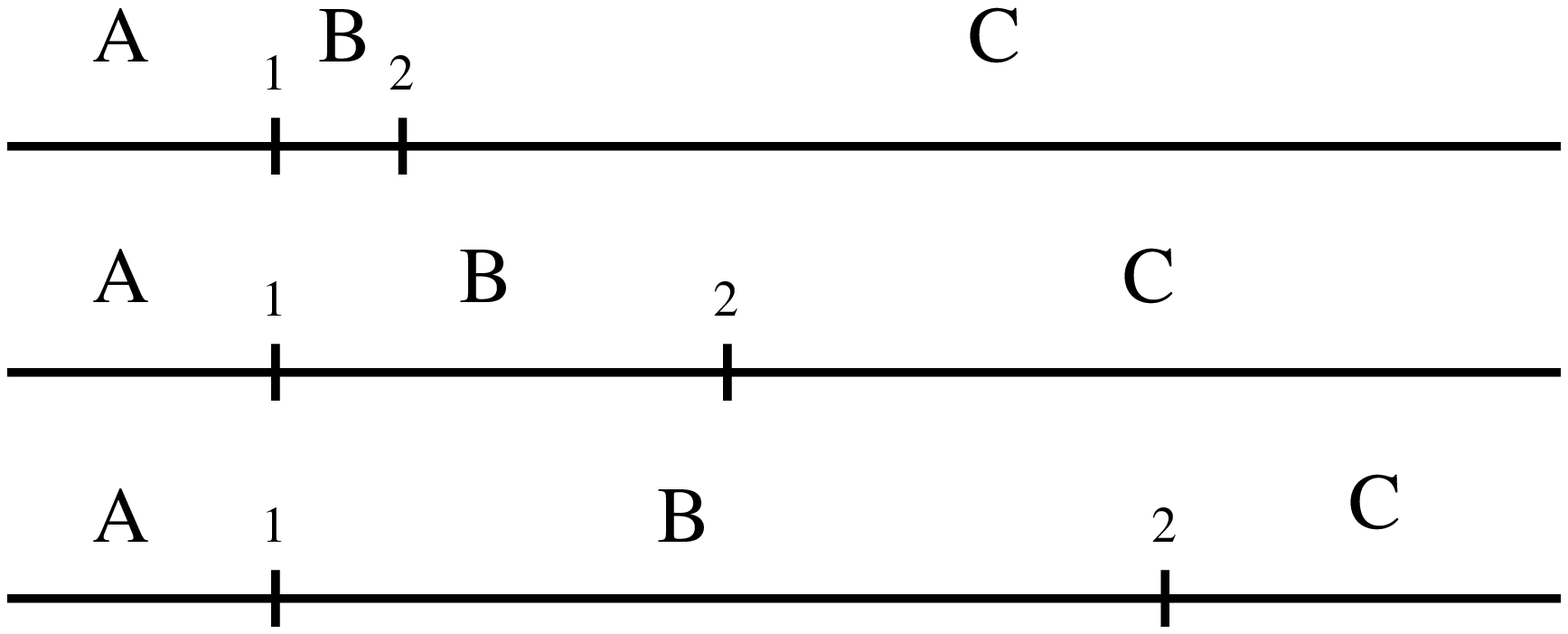}
\caption{\label{sch4}}
\end{center}
\end{figure}

\begin{figure}[ht]
\begin{center}
\includegraphics*[scale=0.6]{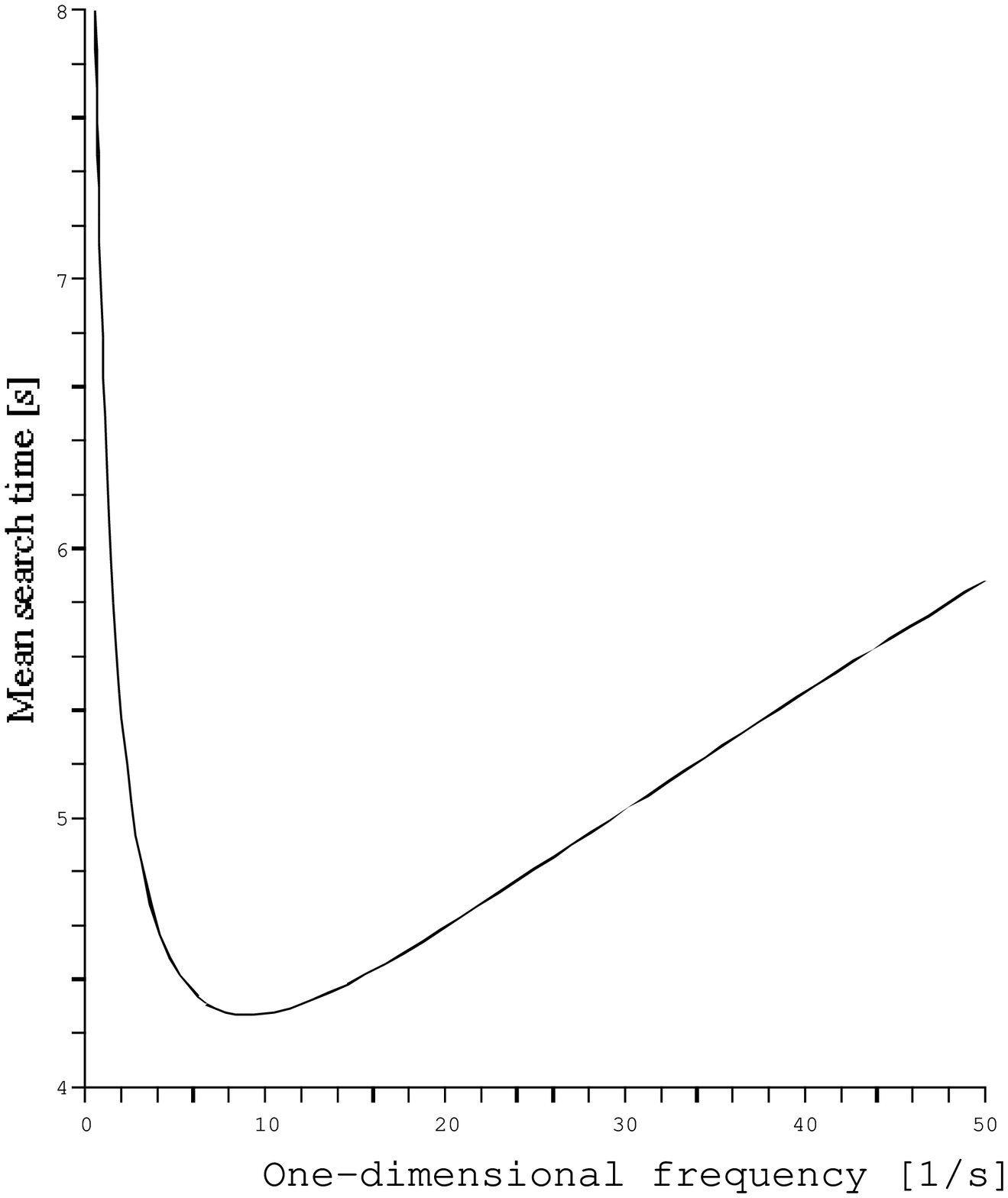}
\caption{\label{Fig1}}
\end{center}
\end{figure}

\begin{figure}[ht]
\begin{center}
\includegraphics*[scale=0.6]{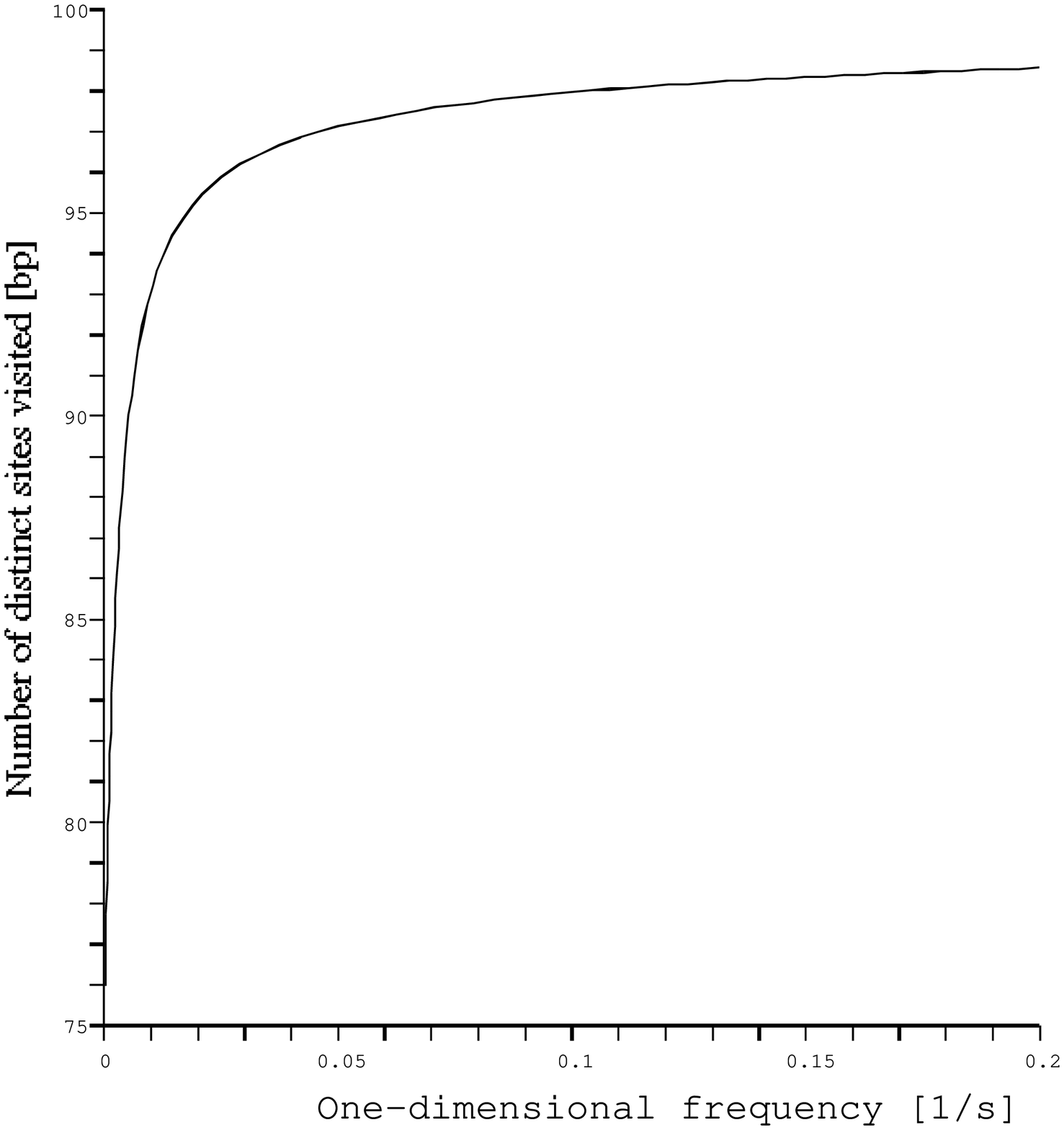}
\caption{\label{Fig2}}
\end{center}
\end{figure}

\begin{figure}[ht]
\begin{center}
\includegraphics*[scale=0.6]{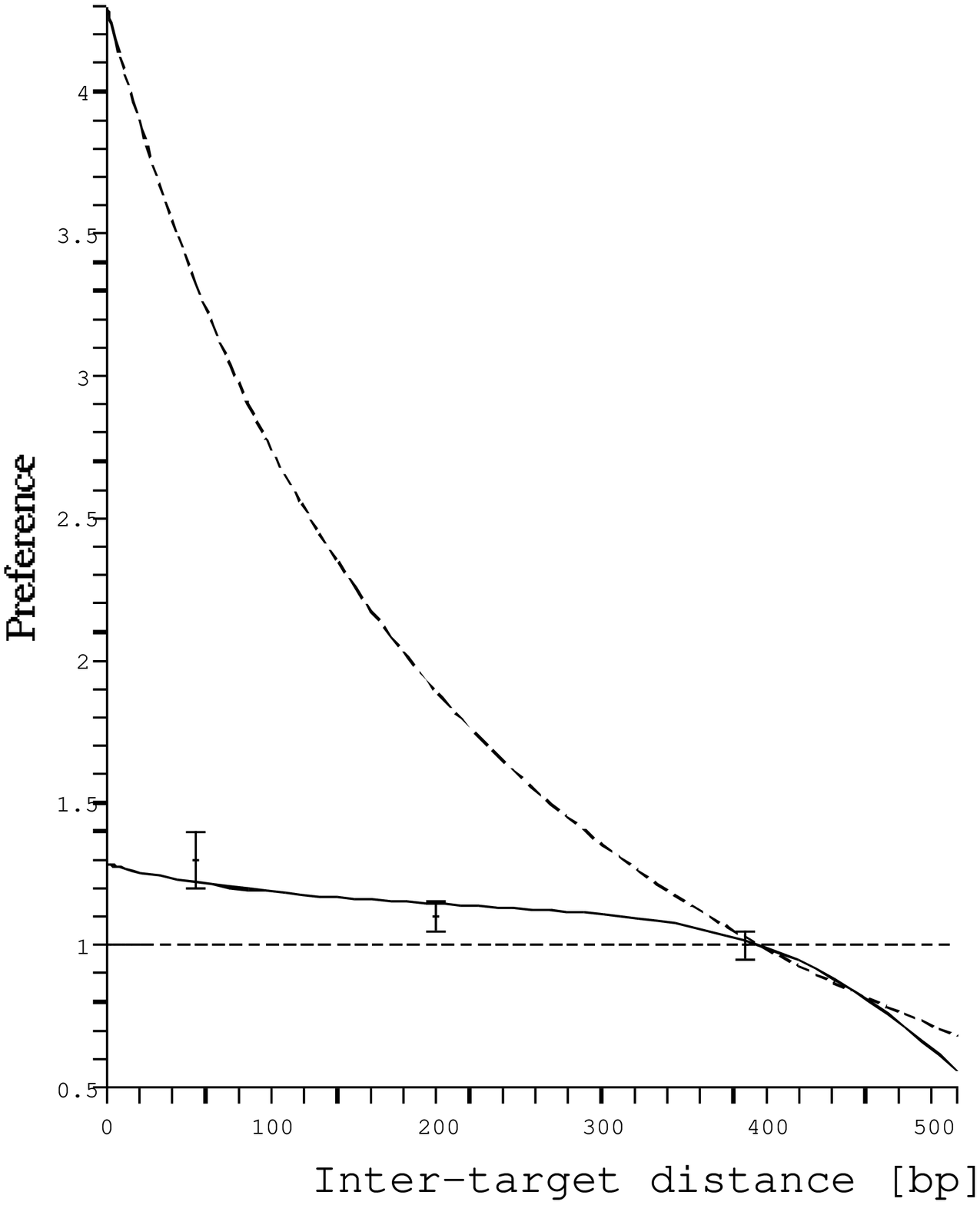}
\caption{\label{Fig3}}
\end{center}
\end{figure}

\begin{figure}[ht]
\begin{center}
\includegraphics*[scale=0.4]{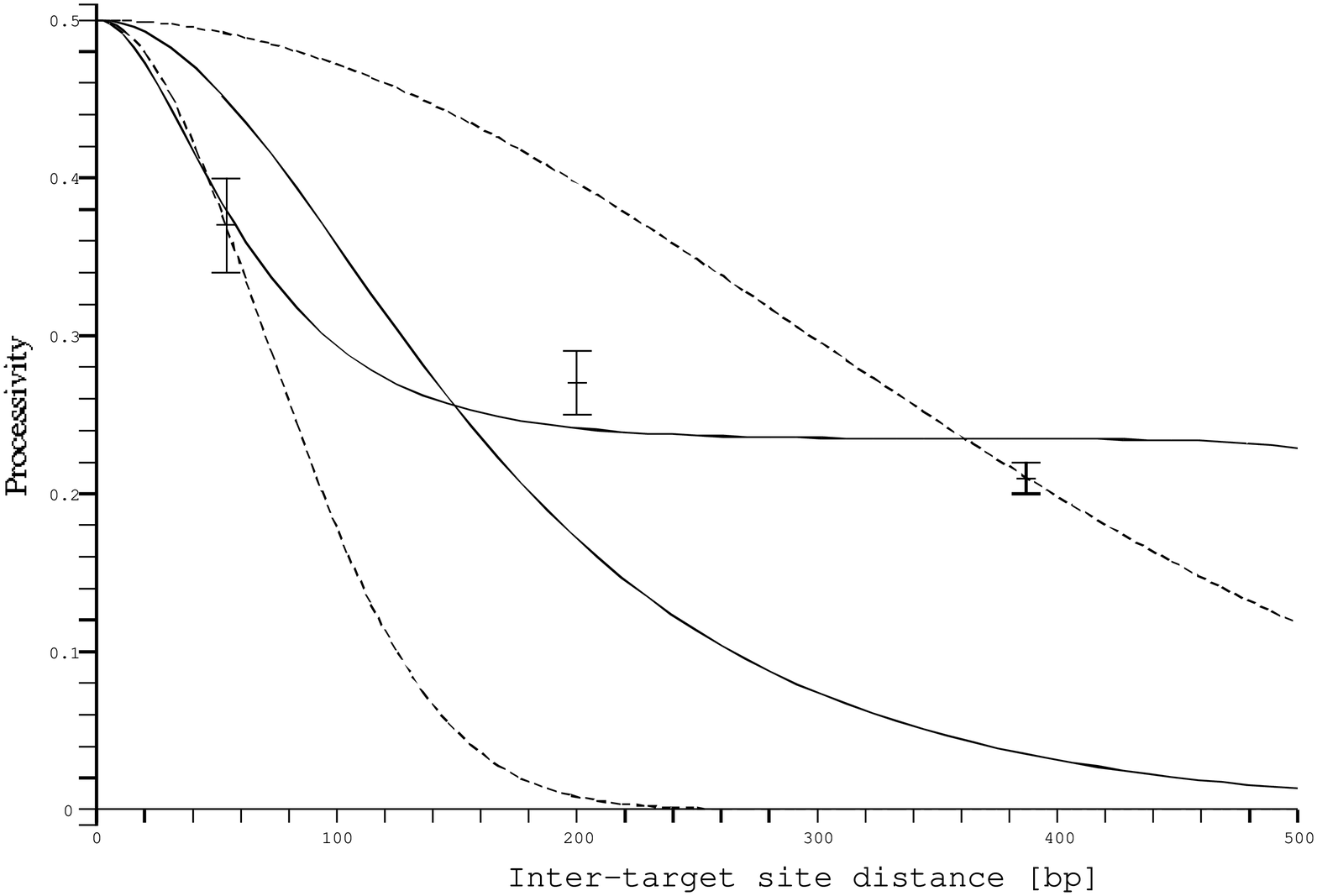}
\caption{\label{Fig4}}
\end{center}
\end{figure}

\end{document}